\documentclass{article}
\usepackage{spconf,amsmath,graphicx,hyperref}
\usepackage{float}
\usepackage{graphicx} 
\usepackage{tcolorbox}
\usepackage{stfloats}  
\usepackage{cuted}
\usepackage{caption}
\usepackage{booktabs}

\title{TART: A Comprehensive Tool for Technique-Aware Audio-to-Tab Guitar Transcription}
\name{%
\begin{tabular}{c}
Akshaj Gupta, Andrea Guzman, Anagha Badriprasad, Hwi Joo Park \\
Upasana Puranik, Robin Netzorg\thanks{Equal Advising}, Jiachen Lian, Gopala Krishna Anumanchipalli
\end{tabular}
}
\address{UC Berkeley}

\begin{document}
\ninept
\maketitle
\begin{abstract}
Automatic Music Transcription (AMT) has advanced significantly for the piano, but transcription for the guitar remains limited due to several key challenges. Existing systems fail to detect and annotate expressive techniques (e.g., slides, bends, percussive hits) and incorrectly map notes to the wrong string and fret combination in the generated tablature. Furthermore, prior models are typically trained on small, isolated datasets, limiting their generalizability to real-world guitar recordings. To overcome these limitations, we propose a four-stage end-to-end pipeline that produces detailed guitar tablature directly from audio. Our system consists of (1) Audio-to-MIDI pitch conversion through a piano transcription model adapted to guitar datasets; (2) MLP-based expressive technique classification; (3) Transformer-based string and fret assignment; and (4) LSTM-based tablature generation. To the best of our knowledge, this framework is the first to generate detailed tablature with accurate fingerings and expressive labels from guitar audio.

\end{abstract}
\section{Introduction}

\begin{figure*}[t]
    \centering
    \includegraphics[width=0.70\linewidth]{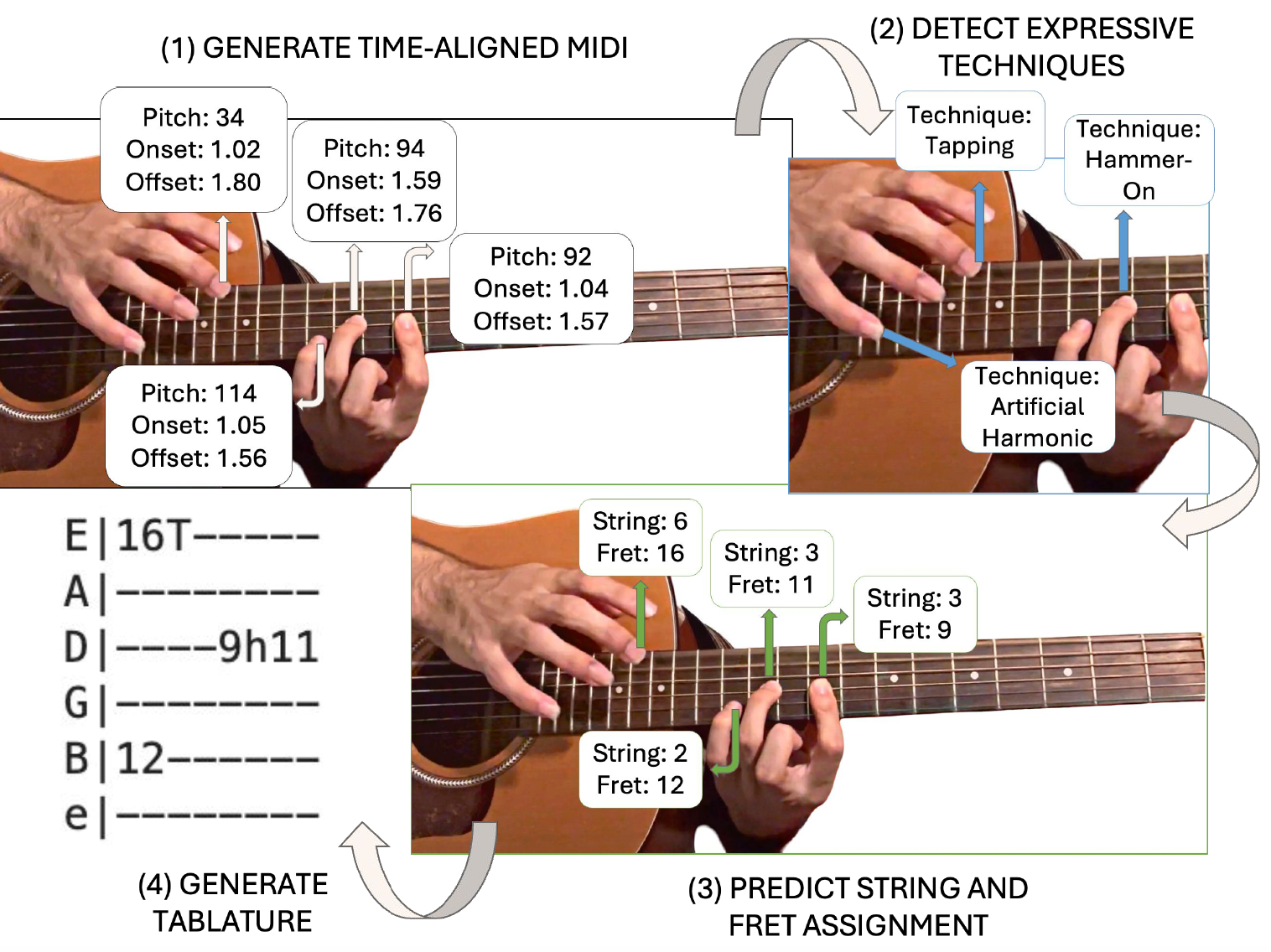}
    \caption{Visual representation of the TART framework. Raw guitar audio is first converted into MIDI note events, capturing pitch, onset, and offset information as shown in stage (1). In stage (2), an expressive technique classifier analyzes the audio to label each note with the corresponding techniques (e.g., hammer-on, tapping). In stage (3), we use a transformer model that takes MIDI note sequences as input and predicts string and fret positions. In stage (4), the gathered data is merged to generate the sample tablature shown.}
    \label{fig:placeholder}
\end{figure*}

Automatic Music Transcription (AMT) is the process of converting an audio recording of an instrumental performance into a standard representation of music, such as sheet music, MIDI, or tablature. AMT has advanced considerably, with piano transcription models achieving impressive note-level accuracy through deep learning architectures such as convolutional and recurrent neural networks, and guitar transcription models following similar architectures \cite{hawthorne-2018-onsets,kong2021highresolutionpianotranscriptionpedals,9222310,9909659, bittner2022lightweightinstrumentagnosticmodelpolyphonic, pedroza2024leveragingrealelectricguitar, li2025pianotranscriptionhierarchicallanguage}. 

In contrast, transcription for expressive musical techniques is lacking. For example, guitarists have developed a wide range of expressive techniques, including percussive hits, slides, bends, and harmonics, which current AMT systems do not capture. Additionally, the guitar is pitch redundant, meaning the same pitch can be produced at multiple string and fret combinations. Current guitar transcription methods typically neglect these nuances, resulting in tablature that fails to represent the expressive and realistic aspects of guitar playing \cite{kong2021highresolutionpianotranscriptionpedals,riley2024highresolutionguitartranscription}. Furthermore, many existing guitar-specific models are trained on limited datasets, which restricts their ability to generalize to real-world performances.

To address these challenges, we propose the Technique-Aware Audio-to-tab Representation Tool (TART), a four-stage transcription framework to generate detailed guitar tablature directly from audio. Our framework builds on a piano transcription model for audio-to-MIDI conversion and further incorporates an expressive technique classifier to capture guitar-specific techniques and a string and fret assignment model to resolve pitch redundancy confusion. Our framework represents one of the first end-to-end approaches to guitar transcription that integrates pitch estimation, expressive technique detection, and string and fret mapping into a unified pipeline.
\section{Relation To Prior Work}
\label{sec:format}

AMT has made significant advancements in piano transcription. A foundational model, Onsets and Frames \cite{hawthorne-2018-onsets}, focused on separately detecting onsets and frame-level pitch events and then combining them for polyphonic transcription. Building on this, Kong et al. \cite{kong2021highresolutionpianotranscriptionpedals} proposed a high-resolution piano transcription model that not only regresses onset and offset times but also models pedal signals, improving overall precision. Extending AMT techniques to the guitar, Riley et al. \cite{riley2024highresolutionguitartranscription} introduced domain adaptation methods to transfer models trained on piano data to guitar datasets. We build on this by developing new adaptation methods and training on larger datasets.

Expressive guitar technique classification is challenging due to the limited data available for training models. Although recent datasets such as Magcil, AGPT, and IDMT-SMT-Guitar~\cite{Mitsou_magcil, Stefani_agpt, kehling_IDMT} increase the diversity of available recordings with annotations for expressive techniques, most datasets remain small.
When it comes to foundational models for classification, Stefani et al.~\cite{Stefani_ExpressiveGuitar-TechniqueClassification} used multilayer feedforward neural networks to train their expressive technique detection model. Our approach further improves on this by testing cross-dataset generalization across domain shifts and addressing the label mismatch between different datasets. 

Traditional pitch estimation methods do not address the fundamental challenge of string–fret assignment on a redundant fretboard. Early approaches cast the problem as constrained optimization over feasible fingerings, such as Sayegh's Optimum Path Paradigm and related dynamic-programming formulations that penalize large positional jumps and enforce simple playability rules \cite{sayegh,radisavljevic}. More recently, researchers have redefined tablature assignment as a translation from symbolic pitch to specific guitar actions. Hamberger et al. introduced the Fretting-Transformer, an encoder-decoder model inspired by the T5 architecture, which effectively maps MIDI tokens to string and fret tokens, achieving state-of-the-art accuracy \cite{hamberger2025frettingtransformerencoderdecodermodelmidi}.

Recent work has highlighted the importance of incorporating expressive technique annotations into playable tablature. Current work involves developing an LSTM-based model to optimize tablature generation and minimize hand jumps while considering technique constraints, extending prior optimization approaches \cite{edwards2024guitartabmml} to produce tablature that is accurate and practical to perform. Other studies have focused on note-level estimation, using sequence models or neural networks to predict onset placement more precisely \cite{wiggins}.

\section{Proposed Framework}
\label{sec:framework}

Our end-to-end tablature transcription framework is composed of three stages that collect data from a guitar audio sample and store them in a unified \texttt{.jams} annotation format and a fourth stage that combines the data to form the final tablature.

In the first stage, Audio-to-MIDI Conversion, we convert the input guitar audio into MIDI notes by predicting the pitch, onset, offset, and velocity for each note. We fine-tune a piano transcription model on guitar data to produce a sequence of notes, along with their timing and pitch attributes, which are stored in a \texttt{.jams} file.

Next, in the Expressive Technique Classification stage, we collect information about the expressive technique being used for each note from a classifier. Each note event in the \texttt{.jams} file is updated with a technique label, enabling the system to represent expressive techniques along with pitch and timing.

In the third stage, String and Fret Assignment, we address pitch redundancy confusion by predicting the most plausible string–fret combination for each note using a transformer-based architecture. These assignments are then appended to each note in the \texttt{.jams}.

Finally, in the Tablature Generation stage, we convert the \texttt{.jams} file into a structured ASCII tablature format. We integrate the note pitch, timing, expressive techniques, and string and fret assignments to produce tablature suitable for performance.

\subsection{Datasets}

Throughout this work, we use the following datasets: 

\label{sec:datasets}
\begin{itemize}
\item \textbf{GuitarSet} \cite{riley2024highresolutionguitartranscription}: High-quality acoustic-guitar recordings with aligned MIDI and note timings across 60 pieces.
\item \textbf{EGDB} \cite{Chen2022TowardsAT}: 240 electric  guitar recordings with MIDI.
\item \textbf{Magcil} \cite{Mitsou_magcil}: 549 electric guitar recordings with technique labels (slides, bends, vibrato, sweep picking, etc.).
\item \textbf{AGPT} \cite{Stefani_agpt}: Acoustic and electric guitar recordings with onset-level technique tags with over 10 hours of audio.
\item \textbf{IDMT-SMT-Chords} \cite{kehling_IDMT}: Note-level technique labels featuring 7,398 chord segments across 4.1 hours of audio.
\item \textbf{SynthTab} \cite{Zang_2024}: 26,181 GuitarPro transcriptions synthetically derived from GuitarPro tracks. 
\item \textbf{DadaGP} \cite{sarmento2021dadagpdatasettokenizedguitarpro}: 15,211 Guitar Pro transcriptions. Able to be converted into MIDI-tab pairs for sequence-sequence tasks. 
\end{itemize}

\section{First Stage: Audio-To-MIDI Conversion}

To train our guitar transcription model, we followed a similar approach of to that of Riley et al. \cite{riley2024highresolutionguitartranscription}, who fine-tuned a piano transcription model built by Kong et al. \cite{kong2021highresolutionpianotranscriptionpedals}. Kong et al.'s model architecture used a convolutional recurrent neural network (CRNN) that takes in log-mel spectrograms and passes them through convolutional layers to extract features, followed by bidirectional GRUs to model temporal structure. We used the same piano transcription model and fine-tuned it on the GuitarSet and EGDB datasets.

Both datasets were processed into a unified HDF5 format containing both audio and MIDI. MIDI files were parsed using the pretty\_midi library to merge the notes into a single channel. We applied an 80/20 train-validation split across all experiments. A combined dataset (GuitarSet-EGDB) was also created by merging the GuitarSet and EGDB datasets.

For all fine-tuning experiments, we split each audio into fixed-length 10-second audio segments with a 1-second hop size and a pitch shift augmentation of $\pm2$ semitones. We used a batch size of 4, a learning rate of $1 \times 10^{-5}$ with a reduction by a factor of $0.9$ every 10{,}000 iterations, and ended after a total of 100{,}000 steps. 

\subsection{Results}

We evaluated model performance during training using four main metrics. Frame average precision (frame\_ap) was used to quantify the model's accuracy in detecting note activations \cite{kong2021highresolutionpianotranscriptionpedals}. Regression onset and offset mean absolute errors (reg\_onset\_mae, reg\_offset\_mae) were used to measure the timing difference between predicted and ground-truth notes. Velocity mean absolute error (velocity\_mae) was used to assess how accurately the model predicts note intensities or loudness. 

For note-level transcription evaluation, we utilize standard measures of precision, recall, and F1 score with a tolerance of \(\pm50\) milliseconds for ground truth versus predicted onset times, called P50, R50, and F50, respectively \cite{riley2024highresolutionguitartranscription}. Here we present the performance results of our guitar transcription framework experiments.

\begin{table}[!htbp]
    \centering
    \caption{Results of onset augmentation experiments}
    \begin{tabular}{cccc}
        \hline
        Onset Aug & Frame\_AP & Onset\_MAE & Offset\_MAE \\
        \hline
        \(\pm10\)ms & 0.9185 & 0.0655 & 0.0887 \\
        \(\pm50\)ms & 0.9197 & 0.0660 & 0.1046 \\
        \(\pm100\)ms & 0.9169 & 0.0655 & 0.1051 \\
        \(\pm200\)ms & 0.9208 & 0.0692 & 0.1088 \\
        \hline
    \end{tabular}
    \label{tab:onset-aug}
\end{table}

Table 1 explores a set of experiments in which we introduced onset augmentation, where each note onset and offset times are randomly shifted by \(\pm10\)ms up to \(\pm200\) ms. This augmentation was intended to improve onset detection but provided minimal benefit. In fact, larger shifts (\(\pm200\)ms) increased error, suggesting that such augmentations introduced noise rather than meaningful variability.

\begin{table}[H]
    \centering
    \caption{Model evaluation on GuitarSet}
    \begin{tabular}{ccccc}
        \hline
        Model & P50 & R50 & F50 \\
        \hline
        Base & 0.731 & 0.705 & 0.704 \\
        Acoustic & 0.839 & 0.841 & 0.837 \\
        Acoustic-Electric & \textbf{0.846} & \textbf{0.836} & \textbf{0.838} \\
        \hline
    \end{tabular}
    \label{tab:guitarset-eval}
\end{table}

\begin{table}[H]
    \centering
    \caption{Model evaluation on EGDB}
    \begin{tabular}{ccccc}
        \hline
        Model & P50 & R50 & F50 \\
        \hline
        Base & 0.543 & 0.710 & 0.596 \\
        Electric & 0.827 & 0.705 & 0.752 \\
        Acoustic-Electric & \textbf{0.793} & \textbf{0.773} & \textbf{0.779} \\
        \hline
    \end{tabular}
    \label{tab:egdb-eval}
\end{table}

\begin{table}[H]
    \centering
    \caption{Model evaluation on GuitarSet-EGDB dataset}
    \begin{tabular}{ccccc}
        \hline
        Model & P50 & R50 & F50 \\
        \hline
        Base & 0.637 & 0.708 & 0.650 \\
        Acoustic-Electric & \textbf{0.820} & \textbf{0.804} & \textbf{0.808} \\
        \hline
    \end{tabular}
    \label{tab:combined-eval}
\end{table}

In Tables 2-4, we evaluated model variants across the datasets. The Acoustic-Electric model was fine-tuned on the GuitarSet-EGDB dataset, the Acoustic model was fine-tuned on GuitarSet only, the Electric model was fine-tuned on EGDB only \cite{riley2024highresolutionguitartranscription, Chen2022TowardsAT}, and the Base model was not fine-tuned on any dataset. The Acoustic-Electric model outperformed the Base, Acoustic, and Electric models on every dataset, demonstrating that combining acoustic and electric audio features enhances transcription accuracy.

\section{Second Stage: Expressive Technique Classification}
\label{sec:print}
In this stage, we decided to classify nine expressive techniques and a "other" class, as detailed in Table~\ref{tab:unified_mapping}.

\subsection{Classification Model Architecture}
\label{ssec:framework}
We employ a regularized feedforward neural network architecture inspired by Stefani et al.~\cite{Stefani_ExpressiveGuitar-TechniqueClassification} as our baseline model. For each detected note onset, we extract a 180-dimensional feature vector (stacking Mel-Frequency Cepstral Coefficients (MFCCs), Bark-Frequency Cepstral Coefficients (BFCCs), mel-spectrogram features, and chroma features), which is processed by four layers of 800 dense units and a batch normalization layer with a softmax output for prediction. To mitigate overfitting, we applied node dropout, L2-regularization, and early stopping during training. 

\subsection{Datasets and Training Methodology}
\label{ssec:datasets}
The datasets used in this phase are the Magcil, AGPT, and IDMT datasets, which collectively focus on acoustic and electric guitars. Upon training on the Magcil dataset ($\sim$500) and testing on the IDMT dataset, we noticed distributional shift and overfitting, despite regularization. To promote generalization, we created a new unified dataset by pooling the three datasets. Sufficient technique-classification accuracy of IDMT after training only on a portion of the IDMT dataset demonstrates justifiability.

\begin{table}[h]
\caption{Expressive Technique Labels Across Datasets}
\label{tab:unified_mapping}
{\scriptsize
\begin{flushleft}
\begin{tabular}{p{1.7cm}p{1.9cm}p{1.7cm}p{1.5cm}}
\hline
\textbf{Unified Label} & \textbf{AGPT Label} & \textbf{IDMT Label} & \textbf{Magcil Label} \\
\hline
picking         & pick over soundhole, pick near bridge & picked           & -- \\
sweep picking   & --                                    & --               & sweep picking \\
alternate picking & --                                  & --               & alt. picking \\
hammer-on, pull-off, legato & --                        & --               & hammer-on, pull-off, legato \\
slides          & --                                    & slide            & slide \\
bends           & --                                    & bending          & bending \\
vibrato         & --                                    & vibrato          & vibrato \\
palm mute       & palm mute                            & palm mute            & -- \\
harmonics       & natural harmonics                     & harmonics        & -- \\
\hline
other           & percussive, etc.                      & dead note, etc.  & tapping, etc. \\
\hline
\end{tabular}
\end{flushleft}
}
\begin{minipage}{.95\columnwidth}
\vspace{1ex}
\scriptsize
\textit{Note:} "--" indicates that the class does not exist in the corresponding dataset. Test results are based on IDMT's seven classes.
\end{minipage}
\end{table}

We preprocessed the data by saving it as file-level annotations. Since AGPT only has onset-start annotations and not duration labels, we used a fixed window of 0.4 seconds as the duration (the median duration value from the IDMT dataset was 0.41s). To show sufficient generalization capabilities of our unified dataset to the IDMT dataset, we trained on 50\% of the IDMT dataset. To balance class distributions, we augmented the data by adding Gaussian noise, temporal, amplitude, and pitch shifts so that each class has at least 2000 samples for training.

\subsection{Results}
\begin{figure}[t]
    \centering
    \includegraphics[width=\linewidth]{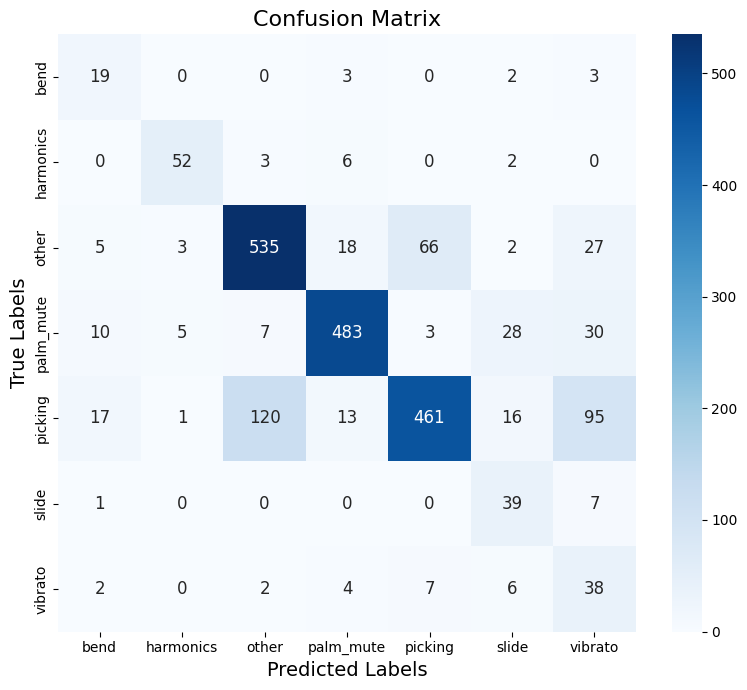}
    \caption{Confusion matrix for the seven test-set-filtered output classes.}
    \label{fig:expressive_technique_confusion_matrix}
\end{figure}

Upon training the unified dataset, we achieved 76\% accuracy on the test set. F1 scores in classes with significantly fewer data points such as bends, slides, and vibrato (each with fewer than 100 data points) were lower, while classes with more support (greater than 500 data points) performed significantly better. See Figure~\ref{fig:expressive_technique_confusion_matrix} for a detailed breakdown.

\section{Third Stage: String and Fret Assignment}
\label{sec:string_fret}

The guitar exhibits pitch redundancy where the same MIDI pitch can be produced at multiple string–fret combinations, creating an optimization problem for tablature generation. To address this, we implement the Fretting-Transformer approach from Hamberger et al. \cite{hamberger2025frettingtransformerencoderdecodermodelmidi}, which treats string and fret assignment as sequence-to-sequence translation using a T5 encoder–decoder architecture with a unified vocabulary for MIDI and tablature tokens. 

\subsection{Dataset Preparation}
We train on symbolic tablature datasets including SynthTab \cite{Zang_2024} and  DadaGP \cite{sarmento2021dadagpdatasettokenizedguitarpro}. We then train on the smaller GuitarSet for fine-tuning to adapt to real recorded guitar performances with MIDI annotations due to its higher quality audio and high-quality annotations.  

    To improve model robustness, we augmented the data with capo positions (0-7). These are represented as conditioning tokens added to sequences, allowing the model to adapt its predictions to different guitar configurations.

\subsection{Evaluation}
Following Hamberger et al. \cite{hamberger2025frettingtransformerencoderdecodermodelmidi}, we evaluate string and fret assignment using pitch accuracy, which ensures the predicted tablature reproduces the correct pitch regardless of fingering, and tab accuracy, which measures whether both the pitch and the specific string–fret assignment are predicted correctly.

To refine predictions, we adopt the post-processing methods from the Fretting-Transformer \cite{hamberger2025frettingtransformerencoderdecodermodelmidi}. First, overlap correction is applied to align temporally adjacent notes. Then, if any predicted string and fret assignment does not yield the correct pitch, a neighbor search is performed. This allows for exact pitch accuracy while providing a slight boost in tablature accuracy and ensuring that outputs remain musically faithful.

\begin{table}[H]
\centering
\caption{Pitch and tab accuracy across acoustic DadaGP}
\small
\setlength{\tabcolsep}{8pt}
\renewcommand{\arraystretch}{1.2}
\begin{tabular}{l c c}
\hline
\textbf{Dataset} & \textbf{Pitch Accuracy} & \textbf{Tab Accuracy} \\
\hline
DadaGP + Postprocessing &94.9\%& 42.1\% \\
\hline
\end{tabular}
\label{tab:dataset-eval}
\end{table}
On the acoustic subset of DadaGP, aligning pitch after inference yielded near-perfect pitch accuracy. With further training on SynthTab and DadaGP followed by fine-tuning on GuitarSet, we expect additional improvements in overall tab accuracy.

\section{Fourth Stage: Tablature Generation}
\label{sec:illust}

After string and fret assignments are determined, we generate human-readable ASCII tablature, aligning fret numbers to strings in temporal order. Expressive techniques are encoded symbolically and appended to fret numbers.

The generator reads notes from the produced \texttt{.jams} file, sorts them by onset time, and fills fixed-width blocks by appending either fret symbols or placeholders to each string line. To validate, we used a synthetic MIDI example with placeholder string and fret assignments and techniques to produce a sample ASCII tablature.

While the rule-based generator helps accomplish our primary goal of providing accurate and realistic transcription, the resulting tablature for complex pieces can be difficult for beginners to play due to large hand jumps or awkward fingerings. To improve playability for beginners, we are also training a LSTM network that simplifies raw guitar pieces into more beginner-friendly versions. We are training on note sequences extracted from MIDI files, where each note is represented by pitch, string, fret, and technique encodings.

We are using a custom loss combining mean-squared error against baseline string and fret mappings with a jump penalty that discourages large changes between consecutive notes. The penalty was scaled to the fretboard distance, ensuring optimization remains physically meaningful. We are also extending the LSTM to optimize under expressive technique constraints, guiding the model to balance technique feasibility with hand comfort.

\section{Conclusion}
\label{sec:foot}

In this work, we propose TART, a four-stage framework for guitar AMT with the goal of resolving the expressive technique detection and string and fret assignment problems. Although our work is still ongoing, our preliminary results are promising, and we will continue to improve our training methods, diversify our data, and gather more results. The long-term objective of this research is to develop a universal music platform for guitarists with many tools including TART.

We plan to test other methods for improving training and final results across our tasks, including evaluating different post-processing algorithms, modifying our model architectures, and training on a significantly larger dataset.

\newpage
\label{sec:refs}

\bibliographystyle{IEEEbib}
\bibliography{strings,refs}

\end{document}